\documentclass[runningheads]{article}
\usepackage{graphicx}
\usepackage[legalpaper, margin=1.25in]{geometry}
\usepackage{authblk}

\usepackage{lineno}

\usepackage{amsfonts}
\usepackage{amsmath}
\usepackage{amssymb}
\usepackage{amsthm}
\usepackage{algorithmic}
\usepackage{algorithm}
\usepackage{bm}
\usepackage{quantikz}
\usepackage{mathastext}
\usepackage{mathtools}
\usepackage{physics}
\usepackage{bbm}
\usepackage[sortcites,sorting=none,style=numeric,backend=biber]{biblatex}
\usepackage{placeins}
\usepackage[hidelinks]{hyperref}
\usepackage[capitalize,noabbrev]{cleveref}

\newcommand{\hperm}{\mathrm{\Pi}}

\newtheorem{defn}{Definition}

\begin{document}
\title{A Mathematical Framework for the Problem of Security for Cognition in Neurotechnology}
%
%
\small
\author[1,2,3]{Bryce Allen Bagley}
\author[1,2]{Claudia Katherina Petritsch}
%
%
\affil[1]{Mathematical Medicine Group, Department of Neurosurgery, Stanford University, Stanford, CA 94305}
\affil[2]{Petritsch Laboratory, Department of Neurosurgery, Stanford University, Stanford, CA 94305}
\affil[3]{Physician-Scientist Training Program, Stanford University, Stanford, CA 94305}
\maketitle              
\begin{abstract}

The rapid advancement in neurotechnology in recent years has created an emerging critical intersection between neurotechnology and security. Implantable devices, non-invasive monitoring, and non-invasive therapies all carry with them the prospect of violating the privacy and autonomy of individuals' cognition. A growing number of scientists and physicians have made calls to address this issue, but applied efforts have been relatively limited. A major barrier hampering scientific and engineering efforts to address these security issues is the lack of a clear means of precisely describing and analyzing relevant problems. In this paper we develop Cognitive Neurosecurity, a mathematical framework which enables such description and analysis by drawing on methods and results from multiple fields. We demonstrate certain statistical properties which have significant implications for Cognitive Neurosecurity, and then describe the algorithmic problems faced by attackers attempting to violate privacy and autonomy, and by defenders attempting to obstruct such attempts.

\end{abstract}

\section{Introduction}


Recent years have seen rapid advancements in both brain-computer interfaces (BCI) and technologies for increasingly precise yet non-invasive readout and alteration of brain function. This progress has been rightfully hailed as the beginning of a revolution in the treatment of countless neurological and psychiatric disorders, and the potential for beneficial impact on human lives is immense.\cite{Alagapan_2023_depression_recovery_dbs,Pandarinath_2017_high_performance_bci_speech_paralyzed_patients,Willett_2021_high_performance_BCI_brain_to_text_communication,Willett_2023_high_performance_speech_neuroprosthesis,Metzger_2023_high_performance_neuroprosthesis_speech_decoding_avatar_control} Even beyond medical applications, there is talk of eventual mass-market use of such technology, and a number of companies are already applying more rudimentary neurotechnology in limited ways for more mundane consumer uses such as entertainment.

These emerging technologies also open the door to malicious actors in profound ways, once conceived of only in science fiction\footnote{As far back as Anthony Gilmore's 1932 novel The Affair of the Brains,\cite{gilmore_2018_affair_of_the_brains}, and perhaps most famously in William Gibson's novel Neuromancer.\cite{Gibson_1984_neuromancer}} yet now rapidly becoming reality.\cite{bahr_2011_cyber_universal_access_security_risks,Landau_EEG_based_bci_security} As neuromonitoring and neuromodulation technologies progress, we must begin to consider the near-inevitability that further abilities to ``read" and ``write" (in some sense) on the human operating system will emerge.\cite{Roelfsema_2018_mind_reading_and_writing_future_of_neurotech,Drew_2023_rise_of_brain_reading_technology,Alagapan_2023_depression_recovery_dbs,Pandarinath_2017_high_performance_bci_speech_paralyzed_patients,Willett_2021_high_performance_BCI_brain_to_text_communication,Willett_2023_high_performance_speech_neuroprosthesis,Metzger_2023_high_performance_neuroprosthesis_speech_decoding_avatar_control} While the level of detail at which this will eventually be feasible remains to be seen, it is already possible to read out thoughts into speech and text with high accuracy at a nearly normal conversational pace,\cite{Willett_2023_high_performance_speech_neuroprosthesis,Metzger_2023_high_performance_neuroprosthesis_speech_decoding_avatar_control} and there is a great deal of active research in increasing the precision of neuromodulation as well. There are also a number of papers which describe experimentally verified means of violating privacy and security and discuss highly plausible risks to autonomy.\cite{bellman_2018_facial_recognition_detection_BCI,Lange_2018_BCI_attack_pin_number_recovery,bernal_2020_brain_implant_neuronal_cyberattacks_experiments_1,bernal_2022_bci_security_review} Historical trends paint a picture of significant risk in the likely outcomes in terms of malicious uses of neurotechnology. Much as the emergence of networking and remote access to computers spurred the development of the mathematical frameworks of Cryptography and Differential Privacy in order to protect data therein,\cite{DuPont_2016_history_of_computer_networks_and_cybersecurity,Dwork_2017_survey_of_reconstruction_attacks_differential_privacy, Vadhan_2017_complexity_of_differential_privacy,Dwork_roth_2014_differential_privacy_textbook,wood_2020_differential_privacy_for_non_technical_audience} we argue that an analogous framework must be developed for those aspects of security and privacy which are unique to neurotechnology. 

To our knowledge the first term proposed for this category of problem was "Neurosecurity", in a 2009 paper by Denning et al.\cite{Denning_2009_neurosecurity_first_paper_neurosurgery_journal} Suggested terms for violations of such security have included "brainjacking"\footnote{We adopt the term "brainjacking" given its emphasis of the fact that attacks on neuro-cyber-physical systems are most analogous to the takeover of other cyber-physical systems, such as drones, self-driving cars, robotic systems, and so on. Such cyberattacks are frequently termed "hijacking", and so our view is that "brainjacking" emphasizes the bio-cybernetic -- in the general sense first expressed by Wiener -- nature of the problem.\cite{wiener_1948_cybernetics_original_book,wiener_1950_cybernetics}} from Pycroft et al and "neural cyberattacks" from Bernal et al.\cite{PYCROFT_2016_brainjacking_bci_security,Bernal_2023_reasons_to_prioritize_BCI_security} A few solutions have been proposed, with the most general thus far likely being  the BCI Anonymizer of Bonaci et al, which would protect privacy by filtering raw neurological data before a BCI is allowed to perform any computations.\cite{bonaci_2014_brain_security_exocortex} A number of papers have provided technical methods, either for specific types of BCI or broader approaches meant to be incorporated at a hardware level from the beginning of BCI design.\cite{bernal_2022_bci_security_review,Belkacem_cybersecurity_P300_BCI,Ajrawi_2021_BCI_cybersecurity_RFID_based_framework} These are promising directions in the form of hardware, though to our knowledge all of these methods require extreme care on the part of manufacturers, and are vulnerable to potential BCI analogs of air-gap attacks\footnote{Air-gap attacks are ones in which even a computer fully isolated from any networks can still be vulnerable to successful cyberattacks. Methods are highly varied and frequently surprisingly creative. Carrara and Adams provide a broad taxonomic survey of such methods.\cite{Carrara_taxonomy_of_air_gap_attacks}} Additionally, a wide variety of potential brainjacking attack types have been described.\cite{bernal_2022_bci_security_review}Prior papers have offered mathematical descriptions of security methods, most notably the work of Bonaci et al which proposes a general description from an information-theoretic perspective.\cite{bonaci_2014_brain_security_exocortex} Because of the precision and fundamental nature of their description from a signals perspective, we construct our definitions of security optimization in \cref{section_readout_attacks} and \cref{section_alteration_attacks} similarly. Indeed, there are natural interfaces between their work and ours, and their BCI Anonymizer is in some ways similar to the noise injection approach we describe in \cref{section_readout_attacks}. 

However, ideally one would like to be able to more directly link cognitive science with the problems of BCI-related security, and theoretical foundations for this have thus far not been provided. Reasons for refining such a link include easier interfacing with psychological experiments, more precise descriptions of the cognitive aspects of BCI security, and the ability to analyze more exotic forms of brainjacking. For example, given BCIs can alter the cognition of individuals, an attacker could potentially avoid detection by individual-centric security approaches by inducing smaller changes in many individuals within a group, lowering their information-theoretic profile at the hardware level and leveraging psychology and sociology to accomplish goals. This is only one example, and we suspect that many types of brainjacking methods targeting multiple individuals and their interactions may be developed in the future. In short, direct hardware security of the various types proposed thus far may not be enough to detect and mitigate all attack types, suggesting that a greater degree of cognitive phenomenology should be incorporated in assessment of various types of attacks.

To this end, in this paper we provide such a cognition-centric approach to BCI security, and term it Cognitive Neurosecurity. This name is chosen over Neurosecurity for two reasons. First, because of our focus on cognition rather than other aspects of brain function. Second, because of the framework's direct applicability to the generally more policy and national security-focused field which shares the name - though BCI security issues are our focus in this paper.\footnote{While the latter will not be addressed in this paper, we are already conducting follow-up work on applying our framework to problems within that field and others.} Drawing on recent literature in mathematical modeling in cognitive science and theoretical neuroscience -- primarily Hyperdimensional Computing and ``Quantum" Cognition -- and concepts from a number of other fields, in this paper we present an analog of Differential Privacy for biological cognition.\footnote{Ienca et al noted this as a possibility in 2018,\cite{Ienca_2018_brain_leaks_cybersecurity_first_mention_of_differential_privacy}, and it has also been briefly discussed at a conceptual level by Bernal et al,\cite{bernal_2022_bci_security_review}, but there does not appear to have been any mathematical analysis of the approach to date.} 

Hyperdimensional Computing (HDC) refers generally to the field which leverages consequences of what has been termed the ``\textit{blessing of dimensionality}" on representation and computation -- particularly in noisy systems.\cite{Kanerva_2009_introduction_to_distributed_representations_hyperdimensional_computing,klyeko_survey_hyperdimensional_computing_part_I,klyeko_survey_hyperdimensional_computing_part_II} That is, as one considers data vectors of increasingly high dimensions, the information content of these vectors becomes increasingly robust to noise \textit{if this information is encoded spatially rather than in more standard forms}. These encodings are termed Vector-Symbolic Architectures (VSA) or Holographic Representations, and represent concepts and their relatedness based on degrees of proximity on the surface of some high-dimensional object.\cite{Kanerva_2009_introduction_to_distributed_representations_hyperdimensional_computing,klyeko_survey_hyperdimensional_computing_part_I,klyeko_survey_hyperdimensional_computing_part_II,Jones_2007_holographic_lexicon_encoding} Traditionally this is a binary hypercube of many thousands of dimensions, but similar properties hold on hyperspheres as well. This will be discussed further in \ref{section_holographic_projective_cognition}. HC has seen success in both theoretical neuroscience and machine learning -- in the latter case particularly in the context of embedded computing.\cite{Neubert_2019_hyperdimensional_computing_for_robotics,Karunaratne_2020_electronics_with_hyperdimensional_computing}

Biological intelligences -- brains, immune systems, and more -- tend to have extremely noisy dynamics and yet retain very robust representations and computations. Unlike in an artificial computer, in a biological one there will not necessarily be one precise value corresponding to ``apple" and another to ``orange", for example. There are known cases of direct, simple mappings from neurons to information, such as grid cells in spatial navigation. Situations of this sort represent what are termed Localist Representations. However, these cells have also been argued to work in concert to represent more complex information.\cite{moser_place_grid_cells_memory_neurobio} Other research has also shown that it is possible in some cases to identify specific neurons which will activate in response to images of only one person.\cite{quiroga_explicit_encoding_single_neurons,gorban_2019_unreasonable_effectiveness_of_small_neural_ensembles} However, even in these cases the simple representations are rendered robust by the high dimensionality of their inputs. In this sense, these small ensembles of cells act as a readout of very high-dimensional data. Additionally, there is good evidence to suggest that neural correlates of many cognitive phenomena require Distributed Representations.\cite{asim_brain_theory_distributed_and_local_both,Kiefer_conceptual_representations_mind_and_brain} In any case, here we are focused on the phenomenology of cognition in our use of HDC, and remain agnostic to the nature of biological representations so as to improve generalizability of our approach. In other words, the results in this paper hold regardless of the specific means by which the brain encodes information biologically. The flexibility of the approach to representation taken in this paper will be elaborated upon in \cref{section_holographic_projective_cognition}. In a classical computer ``apple" and ``orange" would each be associated with some bitstrings, and errors in reading, writing, and other operations could well corrupt the computer's ``understanding" of an object. In other words, a significant fraction of the bits are crucial in aggregate -- even with error-correcting codes. This is akin to how genomes work, but moving to brains and other complex biological computers we find there are cases where individual ``bits" (say, whether or not one neuron fires within a given time window) are in fact not very important, and to a surprising degree.\cite{klyeko_survey_hyperdimensional_computing_part_I,klyeko_survey_hyperdimensional_computing_part_II,asim_brain_theory_distributed_and_local_both,Kiefer_conceptual_representations_mind_and_brain} From the simple observables of human cognition we can see that these neural processes are both computationally sophisticated and based upon a representation which is highly robust. 

There is general agreement that the brain does not operate in a manner any more than very loosely analogous to a von Neumann architecture,\footnote{Overwhelmingly the most common architecture for classical computers.} and some researchers in theoretical neuroscience have argued the brain in fact leverages the statistics of high-dimensional spaces to perform robust computations.\cite{gorban_2019_unreasonable_effectiveness_of_small_neural_ensembles,Gorban_2020_high_dimensional_brain_blessing_of_dimensionality} In this view, brains operate on an aforementioned ``Vector-Symbolic Architecture" (VSA), where vectors in some space encode concepts, with varying degrees of similarity between concepts corresponding to varying degrees of proximity within the space. While VSAs would struggle with noise in a low-dimensional space, as the dimension of a data space increases there is a rapid and nonlinear increase in robustness. 

These high-dimensional spaces have been termed hyperspaces by a community of researchers spanning theoretical neuroscience, computational linguistics, and low-power hardware, robotics, and machine learning, and the hypervectors within them are operated on in what is termed Hyperdimensional Computing (HDC).\footnote{In HDC, hyperspace and hypervector are different from the stricter mathematical definition. While the strict definition simply corresponds to N-spaces of $N\geq4$, in HDC something is considered a hyperspace/hypervector once it approaches or exceeds the order of $10^3$.} hypervectors are generally taken to mean vectors of dimension roughly on the order of $10^3$ or greater,\footnote{It is interesting to note that this is a scale already reachable by at least one commercial quantum computer\cite{Atom_Computing_2023,Ma_2023_atom_computing_1,Evered_2023_atom_computing_2,Scholl_2023_atom_computing_3} and promised by other manufacturers. Though quantum computing is not the focus of this paper, some of its mathematical results can be translated directly to that field.} with increasing dimension offering ever-better performance from a statistical perspective. While there is often discussion in various areas of computer science of the ``problem of dimensionality", HDC leverages an interesting inverse property which has been termed the ``blessing of dimensionality".\cite{brown_1997_original_blessing_of_dimensionality_paper,donoho_summary_paper_blessing_of_dimensionality} That is, as the dimension of a space increases, the distribution of distances between random vectors tightens when normalized by the dimension of the space. Specifically, these inter-vector distances cluster ever more strongly around $0.5$ after dividing distances by the space's dimension. From Brown et al's original use of the term through the two decades since,\cite{brown_1997_original_blessing_of_dimensionality_paper} an increasing range of applications for this property have been identified.\cite{li_2018_embracing_blessing_of_dimensionality,pereda_2018_blessing_of_dimensionality_ADHD_EEG,donoho_summary_paper_blessing_of_dimensionality,Gorban_2018_blessing_of_dimensionality_mathematical_foundations,Gorban_2020_high_dimensional_brain_blessing_of_dimensionality,gorban_2019_unreasonable_effectiveness_of_small_neural_ensembles,liu_2016_blessing_of_dimensionality_mixture_data,anderson_2014_blessing_of_dimensionality_large_gaussian_mixtures,klyeko_survey_hyperdimensional_computing_part_I,klyeko_survey_hyperdimensional_computing_part_II,Neubert_2019_hyperdimensional_computing_for_robotics,Karunaratne_2020_electronics_with_hyperdimensional_computing,kim_hyper_dimensional_complex_vector_low-power_communication,Kanerva_2009_introduction_to_distributed_representations_hyperdimensional_computing} For binary vectors this is an unintuitive yet simple result of the Binomial distribution,\cite{klyeko_survey_hyperdimensional_computing_part_I,klyeko_survey_hyperdimensional_computing_part_II} and in \cref{section_holographic_projective_cognition} we derive similar statistical results for the mathematical structures of Quantum Computing and ``Quantum" Cognition.

``Quantum" Cognition is a model which has gained increasing attention in mathematical psychology and cognitive science in recent decades, and regardless of the perhaps unfortunate choice of name it offers significant advantages over other models of cognition.\cite{Bruza_2015_quantum_cognition_review,Pothos_2022_quantum_cognition_review,Busemeyer_2023_quantum_cognition_models,Yearsley_2016_quantum_cognition_decision_theories,Busemeyer_2015_quantum_cognition_applied_to_psychology,Wang_2013_the_potential_of_quantum_cognition,Wang_2013_quantum_question_order_model,Wang_2014_context_effects_quantum_cognition,Pothos_2009_violations_of_quote_rational_unquote_decisions_quantum_cognition,Trueblood_2012_quantum_cognition_causal_reasoning} It cannot be overly stressed that this approach to cognitive science is in no way related to hypotheses asserting connections between quantum mechanics and consciousness. Instead, quantum cognition simply takes advantage of the fact that the statistical framework based on projective geometry is able to naturally describe the seeming paradoxes that arise from studying cognition. Devised by von Neumann for quantum mechanics,\cite{von_neumann_1955_quantum_probability} the projective framework for probability is an alternative to the more traditional set-theoretic probability of Kolmogorov.\cite{Kolmogorov_1933_foundations_of_probability_theory} Humans are known to exhibit a wide range of behaviors and thought processes which would be paradoxical when assessed through the lens of set-based probability, but which turn out to be captured in a remarkably natural way by projective probability.\cite{Bruza_2015_quantum_cognition_review,Pothos_2022_quantum_cognition_review} To avoid the potential for confusion or misunderstanding in both this paper and future work on Cognitive Neurosecurity, we will use the term Projective Cognition rather than ``Quantum" Cognition, recognizing that the utility of the modeling approach comes in the form of useful features of complex projective geometry.\cite{Bruza_2015_quantum_cognition_review,Pothos_2022_quantum_cognition_review,Busemeyer_2023_quantum_cognition_models} There are a number of these features which have proven an effective and natural fit for cognitive science, including but not limited to:

\begin{itemize}
    \item Contextuality -- the fact that different phrasings of questions or different variations on a situation can produce different results. This is captured by the centrality of measurement operators in determining outcomes in complex projective probability.\cite{Wang_2014_context_effects_quantum_cognition,Bruza_2015_quantum_cognition_review,Pothos_2022_quantum_cognition_review}
    \item Ordering relations -- the phenomenon that the order in which two questions are asked or two events are experienced can alter the probabilities of different responses. This is captured by the non-commutativity of linear algebra.\cite{Wang_2013_quantum_question_order_model,Wang_2013_the_potential_of_quantum_cognition,Bruza_2015_quantum_cognition_review,Pothos_2022_quantum_cognition_review}
    \item Interference effects -- cases where the probabilities of cognitive responses to interacting events violate classical probability. These are captured by superposition.\cite{Pothos_2009_violations_of_quote_rational_unquote_decisions_quantum_cognition,Trueblood_2012_quantum_cognition_causal_reasoning,Pothos_2022_quantum_cognition_review}
\end{itemize}

It should be stressed again that this is not in any way some sort of grand theoretical claim. Rather, it is a simple acknowledgement of three neutral facts. First, that the set-theoretic framework of probabilities in terms of events formulated by Kolmogorov et al -- by far the most widely used one -- is not the only framework.\cite{von_neumann_1955_quantum_probability} Second, that the set/event-centric framework fails to capture many key aspects of human cognition.\cite{Bruza_2015_quantum_cognition_review,Pothos_2022_quantum_cognition_review} Third, that the projective geometry framework developed by von Neumann et al has, by contrast, proven to be a very natural and effective mathematical description of probability in human cognition.\cite{Bruza_2015_quantum_cognition_review,Pothos_2022_quantum_cognition_review} In short, we are simply using a version of statistics which is more effective in the context of cognition. Its historical relationship with quantum mechanics is irrelevant to cognition, just as astrophysics' use of standard probability does not make it relevant to sociology. This paper's use of the non-standard term Projective Cognition reflects a desire to make this clear.

With the interdisciplinary stage of the paper set, we will begin by demonstrating the robustness of a holographic approach to quantum computing and Projective Cognition, as it will serve as a core aspect of our proposed framework for Cognitive Neurosecurity. Inspired by Hyperdimensional Computing and Vector-Symbolic Architectures, we present Projective Holographic Computation as an orthogonal approach to improving the problem of error-tolerance in projective systems like quantum computers (or, for example, AI which could be developed based on Projective Cognition) by leveraging statistical properties of high-dimensional vector spaces. As this paper centers on biological cognition, we focus specifically on one form of Projective Holographic Computation: Projective Holographic Cognition. In the coming sections, we will discuss how it enables analysis of statistical properties of mental states across a range of complexity. The approach is deliberately flexible and robust in its ability to describe not only large numbers of discrete beliefs, preferences, attitudes, etc., but also any number of features of mental states which can be more accurately captured by modeling them as comprised of many individual components.

\section{Methods}

\subsection{Projective Holographic Cognition}\label{section_holographic_projective_cognition}

Projective (or ``Quantum") Cognition typically concerns itself with mental states at the level of complexity accessible to standard experiments in psychology (that is, small numbers of variables for any given context), which is completely reasonable given the historical goals of the mathematical framework.\cite{Bruza_2015_quantum_cognition_review,Pothos_2022_quantum_cognition_review,Busemeyer_2023_quantum_cognition_models,Yearsley_2016_quantum_cognition_decision_theories,Busemeyer_2015_quantum_cognition_applied_to_psychology} However, in order to make our framework of Cognitive Neurosecurity more general,\footnote{``General" in the sense used by physicists, mathematicians, etc., essentially meaning ``capable of describing a wider range of cases".} we must extend it to the higher-dimensional regime. This paper's first focus will thus be on the description of how Projective Cognition extends to high dimensions, and some results on its statistical properties both in terms of internal dynamics of computations and final measurements. We draw on the conceptual origins of Hyperdimensional Computing as a model to explain the robustness of both language and of biological neural computations at the level of large\footnote{On the order of $10^3$ or more.} collections of neurons. The same statistical properties theorized to make neural networks so robust to noise will turn out to have consequences for Cognitive Neurosecurity. We develop this framework in order to facilitate descriptions of the joint effects of the many active cognitive factors which may come into play in a given situation. 

This paper's conception of cognitive states extends that of the existing literature on ``Quantum"/Projective Cognition. In this modeling framework for cognitive science, cognitive state variables are treated as analogous to qubits. This imbues them with inherent contextuality (e.g. the nature of a question or situation impacts how cognition will proceed), interrelatedness (e.g. beliefs about different subjects can interact in a manner which may produce non-classical logic), and other useful features. In modeling cognitive processes we can draw analogies to the evolution of sets of qubits in a quantum computer.\footnote{Again, the analogy to qubits is only a mathematical tool. It in no way suggests or supports claims of a physically quantum basis for cognition.} To the best of our knowledge no such term exists in the literature on ``Quantum"/Projective cognition, and we are extending the model to accommodate arbitrary combinations of beliefs, preferences, and/or any other form of cognitive state variables. For this reason and for the sake of clarity, we will use the term \textit{cogits}  -- a la Descartes' famous ``Cogito ergo sum" -- for these qubit analogues. The intent of cogits is to be a maximally general description capable of integrating any combination of types of cognitive state variables -- and any number of them -- into a single vector description.

For a single instant in time, a state of interest would be a vector of cogits $\psi_t$. In the general case we will assume a high-dimensional $\psi_t$, given the rapid growth in technology for neural measurements and the feasibility of rapidly querying large numbers of cognitive state variables.\footnote{Simply in the sense that people can answer survey questions quickly.} The qualitative differences between low-dimensional and high-dimensional cogit vectors arise from statistical properties of high-dimensional spaces, and are precisely what make a modified version of hyperdimensional computing so useful compared with the standard analysis of states involving only small numbers of cogits.

Because the high-dimensional regime is qualitatively different from the low-dimensional regime, we adopt the novel term Projective Holographic Cognition for discussions of the high-dimensional case. The limited robustness and low representational capacity of small collections of neurons gives way to highly robust and informationally rich representations at higher dimensions, just as the limited content of small numbers of cognitive state variables give way to the richness of human cognition. This paper puts forward two hypotheses regarding this extension to high-dimensional cogit vectors:

\begin{enumerate}
    \item Given projective probability captures the dynamics of human cognition so effectively with small numbers of cogits, including the correlation properties between them, we hypothesize that this efficacy will hold in the high-dimensional regime.
    \item We hypothesize that the statistical properties of cogit holography correspond to the distinguishability of individuals. This will serve as a key source of information in practical applications of Cognitive Neurosecurity both via the structure of correlations between cogits within an individual person and via patterns in these correlations between individual persons at a group or population level. 
\end{enumerate}

Both of these hypotheses are in fact rather tame in that they can be derived from established results. For the first hypothesis, pairwise, triplet-wise, etc. correlations between cogits are known to be effective as a model of interactions among small numbers of cogits -- as verified by extensive work in the field of ``Quantum"/Projective Cognition. Given this, extension to higher dimensions would still hold in at least the limiting case where one only assesses correlations among small n-tuples\footnote{``n-tuples" being the standard term for a collection of \textit{n} things.} of cogits. We are not asserting that these correlations will extend to tuples of larger \textit{n},\footnote{Which to our knowledge have not been studied in the literature on Projective Cognition at the scales discussed in this paper.} but we would not be at all surprised if this holds true empirically. 

The modesty of the second hypothesis will become clearer after our discussion of the statistical properties of cogit holography, and ultimately come down to the blessing of dimensionality. As the dimensionality of a vector space increases, the probability of two vectors being close to one another by chance decreases rapidly.\footnote{Again, normalizing distances by the dimension of the space. This is important to note, as non-normalized distances would inherently increase with the dimension of a space, making the property of declining proximity a trivial one. As a simple analogy to this property, a triangle with two sides of length $1$ would have a hypotenuse of length $\sqrt{2}$, but if we extended to a triangular prism in 3 dimensions it would be $\sqrt{3}$, and so on with higher dimensions.} An analogy in cognition is then the fact that every person exists in a different cognitive state from every other person, and likewise has a mind which works differently from those of every other person. The relevant statistics thus provide a mathematical description for the distinguishability of individuals' cognition. 

\begin{figure}
    \centering
    \includegraphics[width=0.25\textwidth]{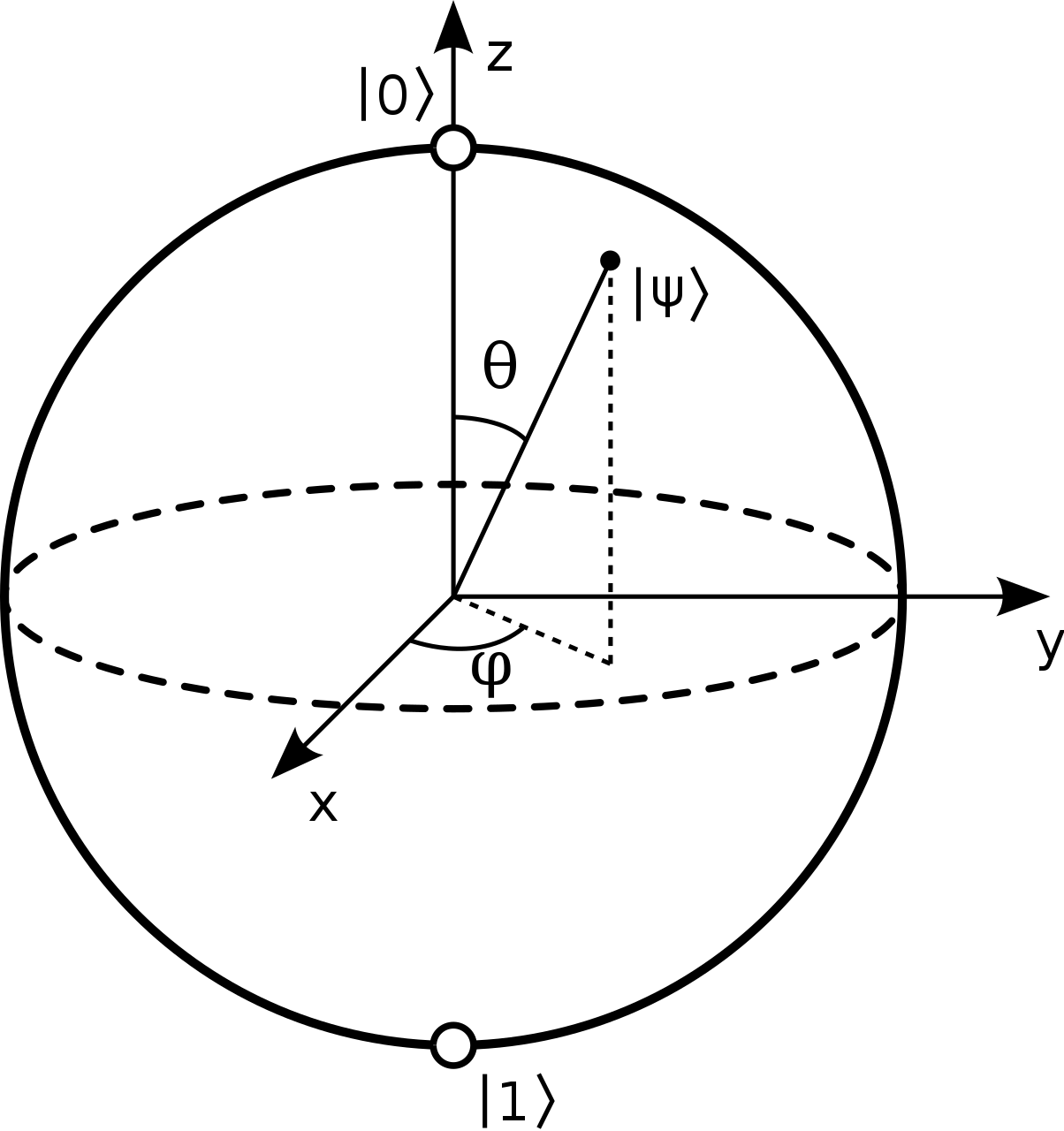}
    \caption{The Bloch sphere graphically represents qubits -- and thus cogits -- in terms of one angle $\varphi$ corresponding to the phasor component and a second angle $\theta$ corresponding to relative probabilities of a qubit/cogit being measured as $0$ or $1$. The notation $\ket{0}$ and $\ket{1}$ are used to represent the states which exist in superposition in a qubit/cogit. Image credit: Wikimedia Commons.\cite{bloch_sphere_wikimedia}}
    \label{fig:bloch_sphere}
\end{figure}

Each cogit can be represented in the same manner as a qubit. Thus a convenient visualization is the Bloch sphere, seen in \ref{fig:bloch_sphere}. In this depiction the vertical axis corresponds to the probabilities of measuring $\ket{0}$ or $\ket{1}$ by the relationship $\cos{\theta}= \text{Pr}[\ket{0}] - \text{Pr}[\ket{1}]$, the difference between the respective probabilities of measuring each of the two outcomes. As one extends this to the multi-cogit case, however, the Bloch sphere ceases to be relevant and different descriptions are needed. This is discussed in the next section, along with Appendix \ref{appendix_holography_statistics}.

\section{Results}

\subsection{Distance Statistics of Cogit Vectors}

In a classical binary HDC, one uses the Hamming distance to describe the similarity of two HV. In other words, taking the absolute difference of the two vectors and summing. Two maximally different HV will have a distance between them equal to the span of the hypercube defining the hyperspece in which those binary HV exist. Taking the same approach for cogits would fail to capture their inherently probabilistic nature, and thus we make use of a more appropriate metric which accounts for features like entanglement and coherence.

A convenient distance metric is the Bures distance, which is derived from fidelity -- itself a measure of the similarity of two states. Here we take two cogit HV $\ket{\psi}$ and $\ket{\varphi}$ sharing the same basis (i.e. corresponding to the same properties), with respective density matrices $\rho$ and $\sigma$. Fidelity is defined as $\text{Fi}(\rho,\sigma) = |\braket{\psi}{\varphi}|^2$ for pure states. Intuitively, this is the simplified case where we assume that we can measure all cogits with any possible relevance. Then the normalized distance metric is yielded by 

\begin{equation}
    \text{b}(\rho,\sigma) = \frac{1}{\sqrt{2}}\text{D}_{\text{Bures}}(\rho,\sigma) = \sqrt{1-\sqrt{\text{Fi}(\rho,\sigma)}}
\end{equation}

This is a normalized version of the Bures distance, and both it and the derivation of relevant probability distributions are discussed in greater detail in \ref{appendix_fidelity_statistics}. One finds that this yields a distribution which decreases exponentially as the normalized distance between two cogit HV decreases. For an $N$-cogit state, the distribution is

\begin{equation}
    \text{Pr}[text{b}(\rho,\sigma) < v] = 1 - \frac{(2v^2-v^4)^{N-1}}{N-1}
\end{equation}

For a distance of even $5\%$ less than the maximum possible distance one already sees the kind of robust concentration of probability which is central to hyperdimensional computing. With $N=100$ there is already a $\sim0.4\%$ chance of even $5\%$ proximity. For a $500$-cogit HV this drops to $\sim2.9\times 1.7^{-5}$. For a distance of less than $50\%$, one has probabilities of $\sim2.9\times 10^{-38}$ and $\sim 1.4\times 10^{-182}$ for $100$-cogit and $500$-cogit states respectively. 

Having discussed a means of linking the mathematics of Hyperdimensional Computing with that of Projective Cognition, we now turn to analyzing the statistical robustness of measurements. In the context of Projective Cognition, these are assessments of cogit states (via questions, behavioral tests, or any other form of assessment).

\subsection{Measurement Statistics}

We have shown that projective hypervector states possess the appropriate distance statistics, and now move to discuss measurement robustness properties. For a given cogit's state $\ket{\psi} = \alpha \ket{0}+\beta\ket{1}$, we have $\text{Pr}(0) = |\alpha|^2$ and $\text{Pr}(1) = |\beta|^2$. For the sake of notational cleanliness in what follows, we will instead use $p=\alpha^2$.

Suppose then that there is some probability $p_i$ for the $i$-th cogit to be measured as state $\ket{1}$ in the absence of noise, and then some uniform probability $q$ for the noise of our measurement process. There are then four outcomes. A ``true" (from the perspective of noise) measurement of $\ket{0}$ with probability $(1-q)p_i$, a true measurement of $\ket{1}$ with probability $(1-q)(1-p_i)$, a false measurement of $\ket{0}$ with probability $q (1-p_i)$, and a false measurement of $\ket{1}$ with probability $q p_i$. Because we are dealing with large numbers of cogits, in the following analysis we can reasonably approximate the $p_i$ as equal, and will use $p$ to denote this. 

This describes a mixture distribution of a noisy measurement of value $x$ with an underlying Bernoulli trial, and has the probability distribution function

\begin{equation*} 
f(x) = (1-q)p^x (1-p)^{1-x} + q(1-p)^x p^{1-x}
\end{equation*}

From here, we construct a description of the distance probabilities for measurements taken of two random projective hypervectors. As with classical hypervector distance statistics, the normalized mean distance would be 0.5. This holds whether the measurement error probabilities are uniform across all cogits -- i.e. for $q_i = q$ $\forall$ $i=\{1,2,...,n\}$ -- or non-uniform, because for random projective hypervectors the cogits will be statistically interchangeable. That is, there would be no significance to the pairing of respective measurement error probabilities with cogits because the indexing of the cogits is irrelevant in describing uniformly randomly distributed projective hypervectors. 

From here, we move to a compound distribution of these mixture distributions, analogous to the move from Bernoulli trial to binomial distribution. In our mixture case, this is described by the function $Pr[k] = \sum_{j=0}^k \left[(1-q)p^j(1-p)^{1-j} + q(1-p)^j p^{1-j} \right]^n$ where $j$ is the number of true Bernoulli ``successes" (in our context, the number of cogits where the two projective hypervectors' measurements would have matched in the absence of error), $k$ is the number of measured ``successes", and $n$ is the number of ``trials" (dimension of each projective hypervector). 

We would like to speak to the case where the measurement errors are not uniform. Using the language of Bernoulli and binomial distributions for clarity's sake, for the Bernoulli/binomial-with-error case we are considering the mean of a single trial $i$ is $\mu_i = (1-q)p +q(1-p) = p + q - 2pq$, and the variance is $\sigma_i^2 = (p+q-2pq)(1-p-q+2pq)$. The mean and variance over $n$ identical trials are $\mu = n\left(p + q - 2pq\right)$ and $\sigma^2 = n(p+q-2pq)(1-p-q+2pq)$. \footnote{Normalizing by the dimension as in prior sections, we have $\mu = p+q-2pq$ and $\sigma^2 = \frac{1}{n}(p+q-2pq)(1-p-q+2pq)$, so again the normalized variance scales down with increasing dimension.}

Applying the central limit theorem to our modified distribution, the probability of two noisy measurements of the same underlying projective hypervector having similarity -- that is, $n$ minus the Hamming distance -- greater than or equal to some $z$ can be approximated by a Gaussian distribution with the preceding mean and variance. We can approximate the cumulative probability function using the Gaussian error function erf by integrating the Gaussian from $-\infty$ to $0$ and then subtracting half the erf function for a given $z\geq 0$. This is equivalent to multiplying the error compliment function, erfc, by $\frac{1}{2}$.

\begin{align*}
    Pr[Z \geq z] & \approx \frac{1}{2}-\frac{1}{\sqrt{2}}\text{erf}\left(\frac{z-n(p+q-2pq)}{\sqrt{2n(p+q-2pq)(1-p-q+2pq)}}\right)\\
    & = \frac{1}{2}- \frac{1}{\sqrt{\pi}}\sum_{i=0}^\infty\left[\frac{(-1)^i}{(2i+1)i!}\left(\frac{z-n(p+q-2pq)}{\sqrt{2n(p+q-2pq)(1-p-q+2pq)}}\right)^{2i+1}\right]
\end{align*}

While $p$ was kept in variable form for the sake of the above analysis, for the statistics of uniformly random projective hypervectors it will simply be $0.5$ as even an error-free measurement would be equally likely to yield 0 or 1. In other words, we can in some sense integrate out the measurement errors when considering random projective hypervectors. Our simplified equation is then

\begin{align*}
    Pr[Z \geq z] & \approx \frac{1}{2}-\frac{1}{\sqrt{2}}\text{erf}\left(\frac{z-n/2}{\sqrt{n/2}}\right)\\
    & = \frac{1}{2}- \frac{1}{\sqrt{\pi}}\sum_{i=0}^\infty\left[\frac{(-1)^i}{(2i+1)i!}\left(\frac{z-n/2}{\sqrt{n/2}}\right)^{2i+1}\right]
\end{align*}

Interestingly, while the measurement errors are very much relevant in the general case, they cancel from the expression when we are considering the distance statistics solely of uniformly random projective hypervectors. The $p=0.5$ case in fact turns out to be equal to the statistics for classical hypervector distances in a setting free from error in measurements of the individual Bernoulli trials. For a 1,000-dimensional projective hypervector distances are clustered tightly around $0.5$, and only about $1$ in $10^{6}$ measurements will have a normalized pairwise distance\footnote{Recall that in HDC one describes normalized distances as the Hamming distance divided by the dimension of the hypervectors.} outside the range $0.425$ and $0.575$.

In fact, the same property holds for assessing fidelity between a true projective hypervector's probabilities and the measured probabilities from repeated tests. In this case we have the measurements as Bernoulli trials of each cogits, with a success simply a question of whether or not a bit-flip error occurred in the process of measuring. The mean is then $n(1-q)$ and variance $n(1-q)q$, and the distances can be described similarly to the above as

\begin{align*}
    Pr[Z \geq z] & \approx \frac{1}{2}-\frac{1}{\sqrt{2}}\text{erf}\left(\frac{z-n(1-q)}{\sqrt{2nq(1-q)}}\right)\\
    & = \frac{1}{2}- \frac{1}{\sqrt{\pi}}\sum_{i=0}^\infty\left[\frac{(-1)^i}{(2i+1)i!}\left(\frac{z-n(1-q)}{\sqrt{2nq(1-q)}}\right)^{2i+1}\right]
\end{align*}

Again using as reference a 1,000-dimensional projective hypervector, with an error probability $q=0.25$ we would find that around $1$ in $10^{6}$ measurements will have a normalized distance from the true measurement that's less than $0.185$ or more than $0.315$. Yet even at an error rate of $q=\frac{1}{3}$ -- with a $1-10^{-6}$ confidence range of $0.26-4.0$ -- we can see that a 1,000-dimensional projective hypervector still allows us to distinguish with high statistical confidence between noisy measurements of the same projective hypervector and measurements of a random projective hypervector. The favorable statistical properties of the geometry of classical hypervectors not only hold in terms of distances between projective hypervector measurements, but also in terms of the difference between projective hypervector measurements and ground truth. The former provides foundational utility, and the latter has has the potential to drastically reduce the number of cognitive/psychological tests and/or sequential runs of an algorithm needed. The blessing of dimensionality thus makes it much easier to confidently distinguish between noise vs. substantive differences in the setting of Projective Holographic Computation -- and thus Projective Holographic Cognition.

As with the classical binary case, the advantage of HDC in the Projective Holographic Cognition setting scales rapidly with increasing dimensionality. For two random hypervectors, as $n$ increases the probability mass of their normalized Hamming distances increasingly concentrates around $0.5$. So too with projective hypervectors. The intuition -- if it can be called that -- is the for any given random vector of ``flipped" bits which moves the measurements away from the probabilities associated with a projective hypervector, the noise will with high probability also move it away from any other random projective hypervector in high-dimensional space. This is the essence of the blessing of dimensionality, and its extension to the case of quantum computing and Projective Holographic Cognition opens the door to potentially radical improvements in the robustness of quantum computation. 

Because there already exists a robust body of literature on algorithms for classical HDC, including previously cited works such as,\cite{klyeko_survey_hyperdimensional_computing_part_I,klyeko_survey_hyperdimensional_computing_part_II,Neubert_2019_hyperdimensional_computing_for_robotics,Karunaratne_2020_electronics_with_hyperdimensional_computing,kim_hyper_dimensional_complex_vector_low-power_communication,Kanerva_2009_introduction_to_distributed_representations_hyperdimensional_computing} but also across the entire field of HDC, this represents a potentially very fruitful intersection and opportunity to adapt new algorithms to quantum computers -- and perhaps to cognitive science as well. As an aside, while we anticipate that HDC's usefulness in quantum computing will depend on context just as it has for classical computing, our hope is that it will not only bring new algorithms to quantum computing but also enable the development of novel quantum algorithms de novo. Reformulating information in terms of distances rather than specific values is a significant departure from the norm in quantum algorithms, but as we have shown, the robustness gains can be substantial. For Cognitive Neurosecurity, however, the robustness of Projective Holographic Cognition is central.

\subsection{Operational Dynamic Analysis}\label{subsection_operational_dynamic_analysis}

We now turn to discussing the practical implications of the statistical properties demonstrated above. Just as in Hyperdimensional Computing, favorable properties of high-dimensional geometry can be expected to make measurements of large numbers of cognitive states remarkably robust. Even with the significant degree of noise and myriad other challenges known to exist in many forms of psychological and cognitive assessment,\cite{Green_2011_psych_testing_assessment_reliability,Meyer_2001_psych_testing_and_assessment_review_of_issues} hypervectors remain easily distinguishable on the scale of such high error rates as to at least mitigate most such effects. Further, correlational structure among cogits in a given projective hypervector would enable error-correction. It is these correlations within individuals' cogits, and those between the cogit hypervectors of different individuals, which are likely to enable greater robustness in machine learning models correlating neural activity with cognition as the scale of cogits considered increases. Additionally, the empirical fact that different people share beliefs in common to differing degrees shows that similarities in the correlation structures of different individuals's cogit hypervectors are all but inevitable. This presents the opportunity for transfer learning from models derived from population-level experimentation over to models which target more specific groups -- or perhaps even individuals. 

Similarly, models of the dynamics of cogit hypervectors over time will almost inevitably have correlations between individuals, because of the empirical fact that some people think more similarly than others. This extends the potential for transfer learning between different models linking neural activity with changes in cogits over time. 

Two general brainjacking scenarios are of interest. The first is attempts by malicious actors to \textit{identify} cognitive states against a subject's will, and the second is attempts to \textit{modify} cognitive states against the subject's will. We term them Readout Attacks (RA) and Alteration Attacks (AA). These could range from blatantly dangerous situations like prisoners being interrogated by an authoritarian regime, to superficially benign yet still fundamentally malign cases like invasive methods of optimizing the manipulative efficacy of advertising or propaganda. Interrogation and focus groups, respectively, taken to new levels. A fine-grained survey of attack types is given by Bernal et al,\cite{bernal_2022_bci_security_review} but for the purposes of this paper we will focus on these two broad categories rather than expanding the discuss to the fine-grained level of differing attack types provided in their excellent overview.\footnote{Essentially all items on the Bernal et al list can be collected under these two broad categories, so follow-up work to this paper could explore with greater depth the application of this paper's framework to the subtypes they present.}

In efforts to read out and/or modify cognitive activity, there are at least three broad levels at which machine learning models could be developed. Because we are not aware of any existing terminology for this topic, the three categories we anticipate are described below. The general goal of analyzing neural activity for cognitive readouts is to connect the dynamics at the relevant scale of monitoring with the cognitive operations they produce. We term this ``Operational Dynamic Analysis", or ODA. The three subtypes we delineate are

\begin{itemize}
    \item $\alpha$-ODA : Methods of analysis which enable the prediction of dynamical correlates of an individual's cognition related to one or more topics, beliefs, cognitive processes, etc. by learning via assessment across a broad population.
    \item $\beta$-ODA : Methods of analysis which enable the learning and prediction of dynamical correlates of an individual's cognition related to one or more topics, beliefs, cognitive processes, etc. via assessment across a group of individuals who are similar according to chosen metric(s) of interest.
    \item $\gamma$-ODA : Methods of analysis which enable the prediction of dynamical correlates of one individual's cognition related to specific topics, beliefs, cognitive processes, etc. by learning via assessment of a single individual.
\end{itemize}

There will presumably be different contexts where each proves most useful, but in general we imagine a progression of model generation and refinement which could move bidirectionally across the range from $\alpha$ to $\gamma$-ODA. Unless neural correlates of belief turn out to have such high variance between individuals that there is a complete lack of consistent patterns for transfer learning, or are so similar that refinement for individual subjects is unnecessary, one could benefit from constructing a general model from a population of subjects. In fact, recent work has shown this exact approach offers significant improvements for the well-studied task of reading out mental states from fMRI scans.\cite{scotti_2024_transfer_learning_MRI_brain_readout} Likewise, one could benefit in broadening or improving the efficacy of a population model via transfer learning from models developed from analyzing single individuals or groups of similar individuals. Given the inherent infeasibility of preventing malicious actors from accessing data from a population sample,\footnote{State actors or other well-funded groups attempting such attacks would have access to the manpower and resources to collect their own data and build their own models.} and all the more so data on small groups or individuals, security considerations will need to center on hampering attacks at the level of individual brains. Acknowledging the impossibility of preventing malicious actors from developing tools, we instead focus on the need for strategic defense methods.

\section{Readout Attacks}\label{section_readout_attacks}

Within the framework of Projective Cognition, for reading out a neural state distribution $Q$ to a cognitive state $\psi$ there will be a learned $\alpha$ model $\{Pr[\psi_t = \varphi_i]\}_{i=1,...,m ;t=1,...,k} = F(Q)$ with $F$ incorporating the $\bm{M}_i$ measurement operators mapping a cognitive state to the probability of some observable $\varphi_i$. That is, a readout of underlying cognition to testable results. For each potential measurement outcome of interest, there can be a different measurement operator.

Using the Probably Approximately Correct (PAC) framework for machine learning, we will have some hypothesis class $F\in H_\alpha$ of potential model functions corresponding to $\alpha$-ODA, and similarly for $\beta$ and $\gamma$. The various $\bm{M}$ within these models will be unitary matrices of the appropriate dimension for $\psi$, as is required to return normalized probabilities. 

A general means of interfering with RA would be to inject noise, much as with the reconstruction attacks of Differential Privacy.\cite{Dwork_roth_2014_differential_privacy_textbook,wood_2020_differential_privacy_for_non_technical_audience} This can be done via the Bonaci et al BCI Anonymizer,\cite{bonaci_2014_brain_security_exocortex} assuming no air-gap attacks are developed for current and future BCI technologies. However, one could also use a secondary implanted system which introduces noise into the neuronal function itself.\footnote{Such a secondary device would need to be physically separate from the main device to at least some extent in order to mitigate air-gap attacks, as magnetic and acoustic air-gap attacks would likely be difficult to defend against without such separation. In this case, the brain's own magnetic activity would serve to shield a physically distanced noise generator from potential air-gapped magnetic attack via the main interface.} Given a sufficiently complex time-varying function for generating a noise kernel and experimentation to optimize such hardware and the noise signals they produce, a physically hardened device would be resistant to both traditional and air-gap attacks while providing a form of privacy much more difficult to circumvent. The goal would be to inject noise such that a metric of model dissimilarity $S$ is maximized over noise which minimizes impact on cognition. For the sake of ease of analysis, we will assume it is technological in origin to improve flexibility in optimizing noise $\aleph$. While we envisage this as described above, the mathematical statements apply equally to something akin to the BCI Anonymizer,\cite{bonaci_2014_brain_security_exocortex}  replacing induced physiologic noise with noise injected between the recording portion of the hardware and the portion performing computations. For AA there would in fact be no difference if it were physiologic noise vs. something like a BCI Anonymizer, but for RA the below would become more similar to the Bonaci et al approach. Our results thus remain applicable even for more indirect approaches to security, in the vein of BCI hardware modifications. However, such approaches retain the greater vulnerability to potential air-gap attacks, and given the stakes and potential gain for malicious actors we expect a great deal of thought and malign innovation will be directed towards developing such attacks as BCI become more common.
 
\subsection{Model-Aware Defense}

We will begin by discussing an approach which assumes some level of knowledge of the possible models attackers may leverage. In a later section we will discuss defense which does not assume any knowledge of the potential models or model types which attackers may employ. However, any bounds on such a Model-Agnostic Defense would almost certainly be sub-optimal for many possible models. To assess Model-Aware Defense one would ideally like to be able to define some analogue of a cryptographic universe. However, the complexity of the Cognitive Neurosecurity problem and breadth of potential elements of $H_A$ and $H_D$ makes constructing such universally strong theorems challenging -- if not impossible. Instead, we acknowledge the limitations of Model-Aware Defense and present relevant ideas before moving to their generalizations in Model-Agnostic Defense.

\begin{defn}[State Model-Optimal Noise]\label{def_state_model_optimal_noise}
    Let $D_{QJS}$ the quantum Jensen-Shannon divergence, $F\in H$ a model mapping distributions of neural activity to distributions of cognitive states, $S$ a normalized metric of distributional dissimilarity in the predictions made by models $F\in H$, appropriate for a given $H$. Assuming a projective cognition model and a classical description of neural activity, for baseline neural state distribution $Q$, neural noise distribution $\aleph$, empiric cognitive noise $\eta(\aleph)$, baseline cognitive distribution $A$, and post-noise cognitive distribution $B = A + \eta(\aleph)$, distribution $\aleph$ is state model-optimal \textbf{if} for a constraining parameter $\lambda$ and a required minimum dissimilarity $\mu \in [0,1]$,
    
    \begin{align*}
        \hat{\aleph} =& \arg\max_\aleph \left( \min_{F\in H} \hspace{1mm}S(F(Q),F(Q+\aleph))\right) - \lambda D_{QJS}(B, -\eta(\aleph)) \\
        & \text{s.t.}\\
        & \min_{F} S(F(Q),F(Q+\aleph)) \geq \mu \hspace{10mm} \forall F\in H
    \end{align*}
\end{defn}

That is, one identifies a noise input distribution which will have the best worst-case performance against all models under consideration, subject to a requirement that prediction must be impaired by a minimum amount. The constraining parameter $\lambda$ allows defenders to weight the relative importance of obfuscation vs. the degree to which protective noise alters cognition.  

The defenders infer $\eta(\aleph)$ from assessment of how different $\aleph$ impact cognition. The quantum Jensen-Shannon divergence (QJSD) is chosen because of its normalization, given the Kullback-Liebler divergence (KLD) is unbounded. We multiply the cognitive noise $\eta$ by $-1$ so that the $D_{QJS}$ quantifies the information needed to identify $A$, as the JSD and its quantum equivalent measure the information needed to identify a mixture of distributions given one of the distributions. In this case, given post-noise observations are drawn from the mixture distribution $B=\frac{1}{2}(A+\eta)$ we want to measure the information needed to identify $A$. This is precisely what the JSD and QJSD measure, as it describes the mutual information between a random variable sampled from a mixture distribution -- true cognitive function plus noise -- and an indicator variable denoting which of the two component distributions produced a given value -- whether it is protective noise or underlying ground truth. With the $-\lambda$ weighting, we seek to minimize this additional information. The goal of an attacker will be to identify a true distribution given observations from some altered distribution, the inefficiency of which is measured by the JSD (and in this case, its quantum analog). In this manner, one aims to leverage the intrinsic noisiness and robustness of biological neural computing in order to confound readouts by producing properties which maximally exploit out-of-distribution weaknesses of readout models, while preserving the underlying cognition itself.

For a population of $n$ subjects, this definition could be modified slightly to additionally take the $\arg \min_{Q\in \{Q_1, Q_2, ..., Q_n\}}$ in order to provide a worst-case bound over some population, or performance could be assessed by averaging over $\{Q_1, Q_2,... Q_n\}$. Similarly, for time-series data $Q_t$ with $t$ representing different times at which neural states are evaluated, and $A_t$ and $B_t$ time series data of the cognitive distributions.

Similarly, one could consider the problem of defending against attempts to characterize the operators describing the dynamics of cognition. In this case, we would have the following:

\begin{defn}[Dynamics Model-Optimal Noise]\label{def_dynamics_model_optimal_noise}
    With $D_{QJS}$ the quantum Jensen-Shannon divergence, $||\cdot ||$ the standard matrix norm\footnote{$\sup_{||v||\neq 0}\frac{||Mv||}{||v||}$}, $G\in H$ a model mapping distributions of neural activity to distributions of operators representing cognitive dynamics, and $H$ a hypothesis class under consideration. Assuming a projective cognition model and a classical description of neural activity, for baseline neural state distribution $Q$, baseline cognitive distribution $A$, and post-noise cognitive distribution $B = A + \eta(\aleph)$, neural noise distribution $\aleph$, distribution $\aleph$ is dynamics model-optimal \textbf{if} for a constraining parameter $\lambda$, and a required minimum dissimilarity $\mu \in [0,1]$,
    
    \begin{align*}
        \hat{\aleph} =& \arg\max_\aleph \left( \min_{G\in H} \hspace{1mm}\frac{1}{2}\mathbf{E}(||G(Q)-G(Q+\aleph)||)\right) - \lambda D_{QJS}(B, -\eta(\aleph)) \\
        & \text{s.t.}\\
        & \min_{G} ||G(Q),G(Q+\aleph)|| \geq \mu \hspace{10mm} \forall G\in H
    \end{align*}
\end{defn}

As with State Model-Optimal Noise, for Dynamics Model-Optimal Noise we can trivially modify the definitions to consider time-series data or population data. Of note, because $G(\cdot)$ is a unitary operator we know $||G(\cdot)||=1$. The operator norm is bounded as $0\leq||G(Q)-G(Q+\aleph)||\leq 2$ - and so $0\leq\mathbf{E}(||G(Q)-G(Q+\aleph))\leq 2$ as well. The division by $2$ ensures that the first term has the same $0\leq \cdot \leq 1$ bounding as the quantum Jensen-Shannon divergence. Population and/or time-series data can be considered as with State Model-Optimal Noise.

Distributions play a key role for both attacker and defender, in this problem. While the attacker will have difficulty predicting the noise $\epsilon$ they will encounter,\footnote{Or at least they must in the case where defense will have any effect.}, without information about the models of attackers, the defenders will themselves have distributions over the respective hypothesis classes. Indeed, the hypothesis classes may be so broad as to be intractable if considered as generally as possible. We can thus take a few different general approaches to optimization of $\epsilon$.

The naive approach is to simply assess a wide ensemble of function classes. Among those with performance above some arbitrary threshold $\mu$, define the union of their function classes as the defender hypothesis class. $H_D = \{\cup_j H_j\}\hspace{1mm}\forall j$ such that $\inf H_j(Q)-H(Q+\aleph)\geq \mu$. In this case, one must make the assumption that the cardinality $|\{\{H_D \oplus H_A\}-H_D\}|$ - in terms of function classes within the attacker and defender hypothesis classes $H_A$ and $H_D$ - is sufficiently small. It is likely impossible to make any precise statements about what would count as ``sufficiently small", given it is impossible to say with certainty what element of $\{\{H_D \oplus H_A\}-H_D\}$ is most robust against the chosen $\epsilon$.

In this approach one then simply performs the optimization problem from \cref{def_state_model_optimal_noise} or \cref{def_dynamics_model_optimal_noise} using the heuristically determined $H_D$. While this is the simplest and most straightforward, the difficulty in making any precise claims about its performance renders absolute security guarantees impossible. Empirical results are straightforward to determine. While not universal guarantees, we expect that such results will be sufficient in many practical settings. In practice cybersecurity requires identification of vulnerabilities and rapid and effective patching of defects which are discovered or exploited. No universal guarantees can be provided for highly complex systems. Similar approach could thus be applied in cases where new model types are discovered, and $\aleph$ adjusted to defend against them. 

To make more general statements, however, we must describe the problem in a way which does not consider specific collections of models.

\subsection{Model-Agnostic Defense}

Any system for an attacker to read out cognitive states will need to use information-theoretic properties of brain activity in order to correlate them with cognition. As laid out very neatly by Bonaci et al,\cite{bonaci_2014_brain_security_exocortex} such information-theoretic descriptions provide a strong foundation for problems of this sort. Here we adapts and extend such ideas so as to interface with the precise mathematical description of cognitive states. This is the space in which $\epsilon$ operates. Meanwhile, we would like to perturb cognitive state $\psi$ as little as possible. This allows us to define optimal noise in a way which is more fundamental, though we do not make conjectures as to whether it or State Model-Optimal Noise (SMON)/Dynamics Model-Optimal Noise (DMON) will be more useful in any given context as there are tradeoffs for each.

The maximal reduction of information in attempts to read out neural activity will be the maximization of the Jensen-Shannon divergence (JSD) of some underlying high-dimensional distribution and a second distribution with added noise. Minimizing the impact on cognition represents a minimization of the the JSD's quantum analog, as in \cref{def_state_model_optimal_noise} and \cref{def_dynamics_model_optimal_noise}. Because these divergence metrics measure the amount of additional information needed to determine one distribution given another (in this case the true distribution given the altered one), these divergences represents exactly the efficacy of readout mitigation. We would like for the adversary to require as much additional information as possible for any given increase in the distinguishability of baseline cognition vs. cognition in the presence of protective noise.

So we then define the following.

\begin{defn}[State Information-Optimal Noise]\label{def_information_optimal_noise}
    Let $D_{J}$ be the classical Jensen-Shannon divergence and $D_{QJS}$ the quantum Jensen-Shannon divergence. Assuming a projective cognition model and a classical description of neural activity, for baseline neural state distribution $Q$, neural noise distribution $\aleph$, empiric cognitive noise distribution $\eta(\aleph)$, baseline cognitive distribution $A$, and post-noise cognitive distribution $B = A + \eta$, distribution $\aleph$ is state information-optimal \textbf{if} for a constraining parameter $\lambda$ and a required minimum dissimilarity $\mu \in [0,1]$,
    
    \begin{align*}
        \hat{\aleph} =& \arg\max_\aleph \left( D_{J}(Q,-\aleph) - \lambda D_{QJS}(B,-\eta(\aleph)) \right)\\
        \text{s.t.}\\
        & D_{J}(Q,Q+\aleph) \geq \mu  \hspace{10mm} \forall F\in H
    \end{align*}
\end{defn}

Similarly,

\begin{defn}[Dynamics Information-Optimal Noise]\label{def_information_optimal_noise}
    Let $D_{J}$ be the classical Jensen-Shannon divergence and $D_{QJS}$ the quantum Jensen-Shannon divergence. Assuming a projective cognition model and a classical description of neural activity, for baseline neural state distribution $Q$, neural noise distribution $\aleph$, function $\kappa(\aleph)$ mapping neural activity noise to cognitive dynamics noise, baseline cognitive dynamics operator distribution $\Lambda$, and post-noise cognitive distribution $\Delta = \Lambda + \kappa(\aleph)$, distribution $\aleph$ is state information-optimal \textbf{if} for a constraining parameter $\lambda$ and a required minimum dissimilarity $\mu \in [0,1]$,
    
    \begin{align*}
        \hat{\aleph} =& \arg\max_\aleph \left( D_{J}(Q,-\aleph) - \lambda D_{QJS}(\Delta,-\kappa(\aleph)) \right)\\
        \text{s.t.}\\
        & D_{J}(Q,Q+\aleph) \geq \mu  \hspace{10mm} 
    \end{align*}
\end{defn}

As with SMON and DMON, one would modify the definition to take the $\arg \min_{Q\in \{Q_1, Q_2, ..., Q_n\}}$ in order to provide a worst-case bound over some population, or average over $\{Q_1, Q_2,... Q_n\}$ if mean performance was the priority. Likewise with time-series $Q_t$, or a combination of the two.

Because this definition is model-agnostic, State Information-Optimal Noise (SION) and Dynamics Information-Optimal Noise (DION) are a more generalized but less targeted counterpart SMON and DMON. The distributions can be defined based on data from a single individual or some collection of individuals, and as long as this is kept consistent it will be able to outperform any given SMON in the general case. Let us now consider the difference between security and similarity. In both model-optimal and information-optimal noise, the first term of the objective function determines security - that is, the difficulty of performing a readout attack - while the second defines dissimilarity between the cognitive distributions with and without noise. 

There is a tradeoff to be had between the two, and neither definition strictly supercedes the other. While SION/DION would offer the broadest protection, SMON/DMON enables a more targeted approach. We anticipate that in at least some contexts SMON/DMON will be able to achieve a given level of protection with notably less alteration of cognitive dynamics than SION/DION. As a simple proof, one can consider a hypothesis class of models which read out the state and/or dynamics from a subset of five variables out of some much larger vector of data.\footnote{The choice of five is arbitrary, and one could equally say ``$y\subset x=\{x_1,...,x_n\}$ with $|y|<<|x|$" to be more general.} While SION would impact all variables, SMON could limit itself to altering only the five assessed by the given models. Although this trivial example is not representative of real settings, it does demonstrate the core principle. For at least some models, SMON/DMON will require less alteration of neural activity than SION/DION for a given level of performance.

\section{Alteration Attacks}\label{section_alteration_attacks}

AA present what may prove to be a more challenging problem. With RA there may be no ``brute-force" approach, so to speak, but for AA a brute-force attack would be both straightforward and difficult (if not impossible) to counter. By increasing activity of reward circuits, one could readily associate some stimulus with positive mental states and in this fashion alter cognition by leveraging baseline neurologic processes. It is likely that defending against AA will require prioritizing detection, and attackers would then face a tradeoff between the degree of alteration and the likihood of their AA being detected.  

Just as with RA, an attacker would aim to construct models from individuals or populations of subjects, and potentially refine them via transfer learning between the $\alpha$, $\beta$, and $\gamma$ models. Beyond that point, however, the problem is the inverse of RA. An attacker would like to maximally alter cognition according to some metric of their choice, and can be assumed to aim to accomplish this with the minimal alteration to neural activity so as to avoid detection. 

For an AA, we thus have definitions corresponding to SMON and DMON, which we term State Model-Optimal Alteration (SMOA) and Dynamics Model-Optimal Alteration (DMOA). There is no analogue to SION/DION, as an attacker will know what model they are using to describe the correspondence between neural activity and cognitive activity.

\begin{defn}[State Model-Optimal Alteration]\label{def_state_model_optimal_alteration}
    With $D_{J}$ the Jensen-Shannon divergence, $D_{QJS}$ the quantum Jensen-Shannon divergence, let $F$ a model mapping from distributions of neural activity to distributions of cognitive states, $Y(\cdot,B)$ a normalized objective function of how close some distribution is to a target cognitive state distribution $B$, and $\aleph$ the distribution of an input signal altering neural activity. For baseline neural state distribution $Q$, baseline cognitive distribution $A \approx F(Q)$, and desired post-alteration cognitive distribution $C \approx F(Q+\aleph)$, alteration signal $\aleph$ is model-alteration optimal \textbf{if} for a constraining parameter $\lambda$
    
    \begin{align*}
        \hat{\kappa} =& \arg\max_\aleph Y(F(Q+\aleph),B) - \lambda D_{J}(Q,Q+\aleph)
    \end{align*}
\end{defn}

This definition is in some sense an inversion of that used by defenders. An attacker seeks to minimize the difference between the distributions of neural states, described by the J-divergence, to avoid detection. Simultaneously, they aim to minimize the difference between the target cognitive state distribution $B$ and the cognitive state distribution inferred via the attacker's model, $F(Q+\aleph)$. In analogy with the defenders' own noise signal, here $\aleph$ is the targeted signal applied by attackers.

Likewise, for dynamics we have

\begin{defn}[Dynamics Model-Optimal Alteration]\label{def_dynamics_model_optimal_noise}
    With $D_{J}$ the Jensen-Shannon divergence, $D_{QJS}$ the quantum Jensen-Shannon divergence, let $G$ a model mapping from distributions of neural activity to distributions of cognitive states, $\Phi$ a target distribution of cognitive dynamics operators, and $\aleph$ the distribution of an input signal altering neural activity. For baseline neural state distribution $Q$, distribution $\aleph$ is dynamics model-optimal \textbf{if} for a constraining parameter $\lambda$
    
    \begin{align*}
        \hat{\aleph} =& \arg\max_\aleph \left( \min_{G\in H} \hspace{1mm}\frac{1}{2}\mathbf{E}(||G(Q+\aleph)-\Phi||)\right) - \lambda D_{J}(Q, Q+\aleph)
    \end{align*}
\end{defn}

These two definitions describe the attackers' efforts to optimize the tradeoff between 1) how closely they approach the desired effect and 2) how readily a defender could detect this alteration attack. The former is described by the first term in the objective functions, and the latter described by $D_{J}(Q,Q+\aleph)$. Minimizing this divergence term reduces the possible information available for defenders to detect.

\section{Conclusions}

In this paper we aimed to initiate an entirely new direction of research. By linking Projective (``Quantum") Cognition -- the unorthodox yet highly effective modeling approach which resolves the major pardoxes of set-theoretic models of cognition -- with ideas from quantum computing, Hyperdimensional Computing, and Differential Privacy, we have provided a framework for precisely analyzing issues related to the fast-emerging problem of Cognitive Neurosecurity. This represents a novel intersection between research on cognitive science, machine learning, privacy and security, artificial intelligence, and neurotechnology, and one which we expect will have significant practical importance in the coming years. The exponential progress of neurotechnology is enabling ever greater medical advancements, but as with all tools it has the potential for both benevolent and malign uses. To the best of our knowledge all existing technical literature has focused on the former, but as has been made clear by the history of computer and data security, if a tool can be used to violate security and privacy for the gain of malicious actors, it \textit{will} be used for precisely that purpose. There is no reason to assume neurotechnology will be any exception, and such a gamble would pose great risk.

Additionally, there are a few interesting notions which arise from the links between fields which have been forged in this paper. First, that quantum computers could be used for simulating cognitive dynamics at scale. This is of particular interest as some devices have reached scales relevant to holographic cognition.\cite{Atom_Computing_2023,Ma_2023_atom_computing_1,Evered_2023_atom_computing_2,Scholl_2023_atom_computing_3} Second, that it may be feasible to develop quantum algorithms or AI systems which interface with human intuition in new, more natural ways than the available classical systems do. Separate from these questions, it is interesting to note that there is nothing requiring the Cognitive Neurosecurity framework be restricted to the individual level. This is not in the same sense as the $\alpha$/$\beta$/$\gamma$-ODA transfer learning described in \cref{subsection_operational_dynamic_analysis}, but in a more general way. Cognitive Neurosecurity is deliberately a general framework to enable greater flexibility, and we suspect it may also prove effective in a somewhat scale-agnostic manner to describe the cognitive dynamics of groups of people -- given the right modifications to the above objective functions. Should BCIs become prevalent beyond the cases of medical necessity for which they are currently used, such coordinated attacks on groups will become increasingly viable and dangerous. Even in the absence of widespread use of neurotechnology, correlates besides neural activity could readily be used for the algorithmic problems described in this paper. In this regard, one could apply the CNS framework to existing forms of cognitive readout such as social media, with the potential advantage over existing methods being that holographic cognition is built on a first-principles empirical description of human cognition. The potential implications and applications of all these possibilities are, however, much beyond the scope of this paper. Additional follow-up work could apply this framework in more specific ways to each of the relevant attack types described in the Bernal et al review.\cite{bernal_2022_bci_security_review}

The question has justifiably been raised of whether it is ethical and safe to put forward the preceding descriptions of the structures of different potential attack types given they also give a mathematical description of a means of accomplishing a broader range of attacks than previously described in the literature.\footnote{Personal communication, Professor Milana Boukhman Trounce, Director of Biosecurity at Stanford University and founding Chair of Biosecurity at the American College of Emergency Physicians.}\cite{bernal_2022_bci_security_review} This is particularly true due to the scale-agnostic  nature of our approach. Indeed, it is likely that some malicious actors will work to leverage these ideas for unethical purposes. But supposing they were not made public, practically speaking it would be impossible to selectively communicate them solely to parties who would use them only for the common good. And in such a case, any attempts at secrecy would be more likely to mainly impact some portion of those who would use it for benevolent ends than those who would use it for malicious ones. As such, we consider the most ethical option to be the release of this paper to the public. While Cognitive Neurosecurity -- being a new area -- is not a traditional domain of Biosecurity, some experts have argued that it will not only very soon become one, but will be a radically new world of opportunities and challenges. Such opportunities and challenges are likely to include some which are analogous to biosecurity problems of other modalities, and many which are wholly unlike anything seen before.\cite{poste_biosecurity_talk_stanford} This is a view we vehemently share.

Finally, there is an interesting notion which emerges from the discussion of attack and defense methods. The Bonaci et al BCI Anonymizer is an innovative approach to the proper design of BCI hardware, but remains vulnerable to air-gap attacks, malicious implants, or BCI hardware vulnerabilities. In order to maximally mitigate such vulnerabilities, one would likely need to induce physiologic noise. This tactic for Cognitive Neurosecurity against readout requires alteration, albeit strategically minimal alteration. And as seen with the similarities between SMON/SION and MOA/IOA, there is a meaningful analogy between AA and defensive measures against RA, and between RA and defensive measures against AA. If no readout is made of cognition and neural activity, then alteration may not in general be reliably detected. The roles of the attacker and the defender exist as odd sorts of mirror images in both cases. While we do not currently see any path towards verifying this one way or another, it leads us to make the following conjecture. There may be no means to achieve physiology-level security guarantees against readout without enacting some degree of alteration to cognition itself, nor such guarantees against alteration without performing some readout. As with much of the myriad implications of this paper and the problems it seeks to begin addressing, there is a great deal of future work to be done on the topic.

\newpage

\large\noindent \textbf{Acknowledgements}\\
\normalsize
We would like to thank a number of people for helping make this paper a reality. Joonhee Choi, PhD, for helpful discussions of quantum computing and the problems of noise in both operations and measurements performed by quantum computers, as these provided helpful background for later work refining the mathematical description of extending ``quantum" cognition to the hyperdimensional regime. Milana Boukhman Trounce, MD, MBA for discussion of this paper from the perspective of biosecurity. Paul J. Wang, MD and Peter Yang, PhD for their support of unorthodox research. Niel Gesundheit, MD, for his encouragement and mentorship. The late Chris Walsh, PhD, for his encouragement, mentorship, and discussions of how thoughtfully-conducted science can help to positively influence the future of humanity's relationship with technology. Sophie Cotton, MS and Sasha Newton for assistance with refining and clarifying the language and phrasing of the paper. Last, but certainly not least, Claudia Petritsch, PhD, for her support of unorthodox research and her mentorship.

\newpage
\printbibliography

\appendix
\section{Hyperdimensional Computing Algebras}\label{section_hyperdimensional_computing_algebras}

In this and the following sections we describe an algebra for HDC on many-cogit systems -- and thus for many-qubit systems as well. We then show that it offers exceptional robustness even on the qubit scales already achievable by existing quantum computers, and equally on the cogit scales which should be feasible to study in applying Cognitive Neurosecurity. The power of HDC comes from both its robustness to noise and its representational richness. While randomly selected points in a hyperspace are not clustered in any way, proportional distances are. This is because of the concentration of probability\footnote{I.E. reduction in the distribution's variance.} with the binomial distribution in bit flips between two vectors, and a brief description of the mathematics behind this will be discussed shortly.

A key challenge in quantum computing is the problem of noise in both the measurements and the computations themselves.\cite{Roffe_2019_quantum_error_correction_review,Corcoles_2020_quantum_computing_challenges_review,Suzuki_2022_quantum_error_correction_review,Resch_2021_benchmarking_quantum_computers_noise} The field thus has the goal of improving the ratio of physical qubits to logical qubits -- the latter being the equivalent number of qubits upon which operations are performed with sufficient robustness. Improvements in error-correcting codes have reduced this requirement significantly, but existing methods still require many physical qubits per logical one, placing significant limitations on what quantum computers are able to accomplish.\cite{deLeon_2021_quantum_computing_hardware_challenges_review,Suzuki_2022_quantum_error_correction_review,sevilla_2020_forecasting_quantum_computing_milestones} Many remarkable innovations have been developed, yet the problem remains a significant practical challenge for applied quantum computing.\cite{Roffe_2019_quantum_error_correction_review,Corcoles_2020_quantum_computing_challenges_review,deLeon_2021_quantum_computing_hardware_challenges_review,Suzuki_2022_quantum_error_correction_review} 

Robustness is a major problem for a number of other areas of computing as well. Examples range from more algorithmic cases like machine learning to more hardware-driven cases such as ultra low-power hardware and biological computing. While not usually considered in the context of quantum computing, biological forms of computing have a long history of inspiring advancements in both machine learning and computer hardware, as detailed in reviews by MacPherson et al, Ullman, and Hassabis et al, among others.\cite{MACPHERSON_2021_neuro_and_AI_history,Ullman_2019_neuro_and_ai_history,Hassabis_2017_neuro_and_ai_history} Recent advancements have included the design of hardware meant to mimic many of the core computational processes of neurons, and their efficiency both in terms of speed and power demands has already garnered much attention.\cite{Boahen_2022_neuromorphic_dendrocentric_learning,Modha_2023_neuromorphic_north_pole,Shastri_2021_neuromorphic_photonic} 

Moving to connect quantum computing with HDC, there are a small number of papers in the area of HDC which reference qubits or quantum concepts, but in each case solely as an analogy and not in a way which is meant to be taken literally. A paper by Bosch et al develops an algorithm for classical hardware which takes the concept of uncertainty before measurement\footnote{In quantum mechanics, there is a fundamental unknowability of any physical value prior to it being measured, and values are thus described by distributions.} and applies it to machine learning. Prior to being measured, the classical state vectors are floating-point valued, and upon ``measurement" they become binary.\cite{bosch_qubit_HDC_not_actually_quantum_just_analogy_used_for_classical_algorithm} Another paper by McDonald et al studied the application of HDC to phasors.\cite{McDonald_phasor_hdc_ai} Yet thus far there has not been any work on applying the principles of HDC -- and the blessing of dimensionality -- to quantum computing. 

\subsection{Why Hyperdimensional Computing?}

We now turn to a brief discussion of the mathematics behind the central advantage of HDC. Using a Hamming distance\footnote{That is, the number of bits in a binary vector which do not match those of a second binary vector.} as our metric on a binary unit $d$-cube and normalizing distances by $d$, vectors are on average $0.5$ apart from each other. In other words, we are defining distance as a fraction of maximum possible distance: the distance between opposite ends of the space. If one performs the calculation, one finds that for a 10,000 bit space the clustering of distances around 0.5 is so tight that we have less than 1/1,000,000 random vector pairs closer than 0.476 or farther than 0.524. This means that the distances between two random vectors are so rarely close as to be negligible. Thus proximity between vectors can represent substantive information with extremely low probability of obfuscation by unrelated vectors in the space. Additionally, the vectors themselves can have a substantial fraction of their bits flipped and still maintain the original distance properties as a consequence of the blessing of dimensionality. Consider three vectors $\Vec{a}$, $\Vec{b}$, and $\Vec{c}$. In a hyperspace, if $\Vec{c}$ starts close to $\Vec{a}$ but moves away from it along some set of dimensions, it is with high probability also moving away from $\Vec{b}$. The key point is that this property makes hyperdimensional representations extremely robust to noise.

As discussed in the introduction, mathematically Localist Representations use positions of vectors in a space to define objects regardless of distances between the vectors. By contrast, in a Vector-Symbolic Architecture -- a.k.a. Distributed or Holographic Representation -- the specific positions of vectors do not convey information in the absence of their context like they would in a Localist Representation.

Whereas with Localist Representations we must add additional dimensions to cope with additional complexity or a greater number of objects, HDC begins with a fixed, high dimension and one simply adds additional hypervectors for new concepts. It is overwhelmingly likely that these new hypervectors will be quasi-orthogonal.\footnote{Meaning that the angle between two vectors is within a small range around $90^\circ$.} That is, the angle between any two hypervectors will with very high probability be very close to $90^\circ$. This means that we do not need to impose any particular restrictions on their placement unless the number of concepts to be encoded by hypervectors is quite large compared with the dimension of the space. The high probability of quasi-orthogonality between random hypervectors is a more precise mathematical statement of the blessing of dimensionality. If we wish to encode a concept which is similar to some others, we may enforce that its distance(s) to one or more other vectors fall within an appropriate region of the space by constructing a random vector whose angles with respect to the others are within some limit of $0^\circ$. 

HDC has a number of variations, but all of them are fundamentally algebras comprised of hypervectors and certain required operations.\footnote{For those unfamiliar with the general theory of algebras: Ordinary algebra and linear algebra are the two most widely known examples of algebras, but many more are possible.} Supposing a set of random binary vectors in a $d$-dimensional space, it is easy to show that vectors will, on average, differ by half of their bits and thus have an average hamming distance of $d/2$ from any other random vector.\footnote{For each bit, two random vectors have probability $0.5$ of being different. Summing over the dimension of the hypervector, we have $d/2$.} However, while the mean remains unchanged, the variance of the distribution becomes increasingly favorable. As discussed in \cref{section_hyperdimensional_computing_algebras}, this same property holds for hypervectors of qubits -- and thus of cogits as well -- just as it does for binary hypervectors.

All forms of HDC represent algebras, but there is flexibility in precisely \textit{what} algebra a given HDC framework will be. Some key requirements are as follows.\footnote{A number of review articles discuss this, but here we make use of the very clear summary provided in McDonald et al's discussion of phasor-based HDC.\cite{McDonald_phasor_hdc_AI} This is particularly helpful given they construct a complex-valued holography which is helpful for representing hyperdimensional collections of cogits.}

\subsection{Required Features}

In addition to the operations of the algebra, any HDC algebra must contain a similarity measure. In binary hypervectors this is the Hamming distance, which is equivalently the cosine similarity for the binary hypercube.

\begin{figure}
    \centering\hspace{-17mm}
    \includegraphics[width=0.85\textwidth]{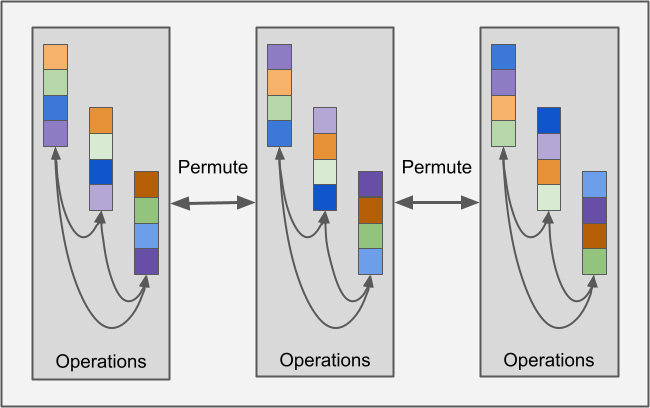}
    \caption{Visual summary of the general structure of a Hyperdimensional Computing algebra. Colored boxes correspond to individual variables, with different shades -- of orange, green, blue, and purple respectively -- denoting different values. The addition, binding, and dynamics operations convert between individual hypervector. Each darker grey box represents a particular pairing of indices with variables, and permutation rotates these pairings in the manner shown.}
    \label{fig:quho_category_summary}
\end{figure}

\subsubsection{Holographic Addition}

For a maximally rich holography, we require that our addition allow the identification -- at least approximately -- of which vectors were added.

Supposing a dictionary of symbols $\{a,b,c,d\}$, for an addition operation $+$, we have

\begin{equation}
    s = [a+b+c]
\end{equation}

Where it must be the case that

\begin{equation}
    sim(s,a) \sim sim(s,b) \sim sim(s,c) >> sim(s,d)
\end{equation}

This means that if we know the dictionary/dictionary of symbols, we can construct the set summed over based on which vectors it is most similar to. The sharp clustering of $sim(\cdot,\cdot)$ around $0.5$ for random hypervectors makes this a robust pseudo-inversion (for lack of a better term), rather than a traditional algebraic inverse. 

\subsubsection{Holographic Binding}

The multiplication operator in a holographic representation is termed ``binding". In contrast with addition, the output is not similar to the inputs.\footnote{In terms of vector similarity. The algebra remains closed under binding just as under addition.} However, while the reversibility via similarity is not present, depending on the specific HDC algebra reversibility may still be provided via other means. 

For cases where binding is self-reversing, we have the following useful property:

\begin{align*}
    s =& [a\otimes b + c\otimes d]\\
    a\otimes s =& a\otimes [a\otimes b + c\otimes d]\\
    =& a\otimes a\otimes b + a\otimes c\otimes d\\
    =& b + \eta\\
    =& \sim b
\end{align*}

While reversibility of binding is not required for an HDC algebra, self-reversibility or the existence of inverse elements can be found in some versions of them. Since an arbitrary hypervector combination is nearly always going to be pseudo-orthogonal, whether $\eta$ is the result of a self-reversal or binding with an inverse element, it will for all intents and purposes be random noise compared with our symbol dictionary. Thus, within the context of a symbol dictionary we can robustly invert binding so long as the inverse (or binding operation itself, for a self-inverse) can recover with high probability only one element of a symbol dictionary. That is, recover one region of the hyperspace which corresponds to a specific symbol.

\subsubsection{Holographic Permutation}

Unlike most typically used algebras, holographic representations come equipped with a third operation: permutation. Permutation here is not the randomized permutations of combinatorics, but rather a rotation in the indices. 

The standard notation for permutation is $\hperm_j$, with $j$ the rotation count. In other words, $\hperm_j$ means that the $0$-th index values for all hypervectors have been shifted to the $j$-th index, the $1$st index is now the $j+1$-th, the $d$-th index now the $j-1$-th, and so on. Holographic permutation is thus also invertible, simply by using $\hperm_{-j}$ or equivalently $\hperm_{d+1}$. 

A key value of this is that it can enforce ordering. We can prevent commutativity and associativity of the binding operation specifically where desired, simply via permuting in between sequential bindings. For example, 

\begin{equation*}
    \hperm_3 c \otimes \hperm_2 b \otimes \hperm_1 a \neq \hperm_2 c \otimes \hperm_1 b \otimes \hperm_3 a
\end{equation*}

\begin{align*}
    (a\otimes b)\otimes(c\otimes d) &= (a\otimes d)\otimes(b\otimes c)\\
    (a\otimes b)\otimes \hperm_1(c\otimes d) &\neq (a\otimes d)\otimes (b\otimes c)
\end{align*}

\section{Complex Holographic Algebra}

The following summary of a basic Complex Holographic Algebra is drawn from a paper by McDonald et al, who are to our knowledge the first to develop one in the literature.\cite{McDonald_phasor_hdc_ai} In a complex holographic algebra the symbol universe is composed of hypervectors not in $\mathbb{R}^d$, but $\mathbb{C}^d$. From this universe, some collection of regions correspond to different symbols within our dictionary. 

In discrete time this would take the form $e^{i2\pi(t/r)}$ with discrete time $t$ and period resolution $r$. In continuous time, we can of course simply rescale $t$ however we like.

\subsection{Similarity in a Complex Holographic Dictionary}

Just as with real-valued hypervectors, cosine similarity is the metric we can use for complex hypervectors.\footnote{Recall that binary hypervectors use Hamming distance, which is the binary analog of cosine similarity. Both will be used for Quantum Holography.} 

\begin{equation}
    sim(\psi,\varphi) = cos(\psi,\varphi) = \frac{\psi \cdot \varphi}{||\psi||||\varphi||} = \frac{\sum_i \psi_i \varphi_i}{\sqrt{\sum_j \psi_j \psi_j}\sqrt{\sum_j \varphi_j \varphi_j}}
\end{equation}

\subsection{Complex Holographic Addition}

The most straightforward version of addition is simple complex number addition, but with a further normalization to retain a unit-magnitude phasor. This results in a loss of some information, making complex hypervector addition only approximately reversible. 

\subsection{Complex Holographic Binding}

Binding in the complex holographic algebra is the same as a normal product of two phasors, so we simply add the exponents. 

\begin{equation}
    \psi\otimes \varphi = e^{i\theta_\psi} \otimes e^{i\theta_\varphi} = e^{i(\theta_\psi+\theta_\varphi)}
\end{equation}

Inversion then becomes multiplication with the complex conjugate, 

\begin{equation}
    \psi^*\otimes \psi \otimes \varphi = \varphi
\end{equation}

\section{Operations for Projective Holographic Algebra}

The algebra for cogits will contain the three essential operations of all hyperdimensional computing algebras, along with an additional two operators for measurement and dynamics. The first is added because for both qubits and cogits one must make phenomenological measurements and cannot assess certain properties of a state without measuring it. The second is added because binding of hypervectors is insufficient to describe the fact that underlying processes transform the hypervectors without needing to interact with another hypervector. 

\subsection{Similarity}

We are not aware of any mechanism for assessing the similarity of qubit or cogit hypervectors prior to measurement. However, describing measurements is very straightforward, and also sufficient for us to use the advantageous properties of HDC.

\subsection{Addition}

For projective hypervectors, superposition offers a very natural way to perform hypervector addition. Because superposition is not invertible, per se, PHA lacks an additive inverse.

\subsection{Binding}

Here we can handle phase binding via the mathematical formulation of what are in quantum computing termed Controlled-Phase gates, which will add the phase angles of two cogits. By matching cogits pairwise from two projective hypervectors, this extends directly to the many-cogit setting. 

\subsection{Measurement}

Measurement for the PHA is simply the normal measurement process of projective statistics. For some state $\ket{\psi}$ and state of interest $\ket{a}$ one has the standard Born rule:

\begin{equation*}
    Pr[a] = \braket{a}{\psi}
\end{equation*}

Or, if one prefers to describe measurements in terms of measurement operators with a set of $M_a$ corresponding to each state of interest $\ket{a}$:

\begin{equation*}
    Pr[a] = \bra{\psi}M_a \ket{\psi}
\end{equation*}

\subsubsection{Dynamics}

Once again using the standard mathematical formalism, dynamical processes transforming one state $\psi$ into another state $\psi'$ are described as 

\begin{equation*}
    \ket{\psi'} = H\ket{\psi}
\end{equation*}

where $H$ is the operator describing the dynamics of the system.

\subsection{Permutation}

Because cogits share their basic mathematical properties with qubits, they similarly require a vector of length $n=2^N$ to describe $N$ cogits. This is because the hypervector is formed via taking the tensor product $\ket{\psi} = \ket{\psi_1}\otimes\ket{\psi_2}\otimes\cdots\ket{\psi_{N-1}}\ket{\psi_N}$. Similarly, if an $x$-permutation $\ket{\psi_i} \rightarrow \ket{\psi_j}$ such that $i=\{1,...,N\}\rightarrow j=\{N+x, 1,...,N-x-1, N-x\}$ is performed, any operator $H$ describing dynamics of $\ket{\psi}$ must be rearranged according to that same change in ordering of the hypervector.

\section{Statiscital Properties of Projective Holography}\label{appendix_holography_statistics}

While the phasor-based HDC algebra developed by McDonald et al. is effective for complex-valued data such as classical analog signals processing, for example,\cite{McDonald_phasor_hdc_ai} it is insufficient for our purposes. With qubits we must additionally account for whatever property (often spin) is to be mapped to measurement outcomes -- and likewise for cogits. The Bloch sphere\footnote{A graphical representation of a qubit, where each axis corresponds to one of the three Pauli matrices whose weighted sum (normalized to unit length) describes a qubit.} provides a convenient representation for this, and is shown in \cref{fig:bloch_sphere}. Describing the statistics of distances between purely random cogits could be quite problematic in some approaches to formulating the problem, and while distances are uniformly distributed if assessed in terms of angular distances rather than surface distances assessing them in those terms fails to account for entanglement and other relevant properties.

\subsection{fidelity Statistics for Pure States}\label{appendix_fidelity_statistics}

A useful simplification is to assume that the set of cogits being considered is a comprehensive description of an individual's cognitive state. This is equivalent to a pure state in quantum mechanics, and assumes there are no correlations with cogits outside those being directly studied. 

State to state fidelity $\text{Fi}(\rho,\sigma)$ is the most straightforward function to use. In order to have an analog of the $0 \leq d \leq 1$ normalization from the hypercube distances between binary classical HV, we specifically use a rescaled Bures distance. Specifically, the standard Bures distance rescaled by $1/\sqrt{2}$.

\begin{equation}\label{equation_normalized_bures_distance}
    b(\rho,\sigma) = \frac{1}{\sqrt{2}}\text{D}_{\text{Bures}}(\rho,\sigma) = \sqrt{1-\sqrt{\text{Fi}(\rho,\sigma)}}
\end{equation}

Given states $\ket{\psi}$ and $\ket{\varphi}$ with corresponding density matrices $\rho = \ket{\psi}\bra{\psi}$ and $\sigma = \ket{\varphi}\bra{\varphi}$, for pure states (which is what we are considering in this simplified case) fidelity is defined as

\begin{equation}
    \text{Fi}(\rho, \sigma) = |\bra{\psi}\ket{\varphi}|^2
\end{equation}

Without loss of generality, we can transform both states so that $\ket{\psi}$ aligns with the standard basis vector $(1,0,...,0)$. The uniform distribution over the complex hypersphere is preserved under unitary transform, so the fidelity distribution simplifies to the distribution of the squared magnitude of $|\tilde{\varphi_1}|^2$, the first element of the other post-transform state -- which is identically distributed to all other $|\tilde{\varphi}_i|^2$.

Next we turn to consider concentration of fidelity. Because $|\braket{\varphi}{\varphi}|^2 = 1$ and the elements are drawn randomly from the unit complex hypersphere, the $|\varphi_i|^2$ are drawn from the symmetric Dirichlet distribution with parameters $\alpha=1$ and $K=N$. This means that $\mathbb{E}[|\varphi_i|^2] = \mathbb{E}[\text{Fi}(\rho,\sigma)] = 1/N$. Equivalently, by marginalizing over all contributions except $|\tilde{\varphi}_i|^2$, we have the distribution $\text{Beta}(1,N-1)$. Thus for some fidelity value $y$, the probability of having lower fidelity is

\begin{equation}
\begin{aligned}
    \text{Pr}[\text{Fi}(\rho,\sigma)<y] = \int_0^y dx (1-x)^{N-2} = & \frac{(1-y)^N}{(N-1)(y-1)}- \frac{1}{(N-1)(-1)} \\
    = & \frac{1}{(N-1)} - \frac{(1-y)^N}{(N-1)(1-y)} \\
    = & \frac{(1-y)^{N-1}}{N-1}\\
    = & \frac{\left(1-|\braket{\psi}{\varphi}|^2 \right)^{N-1}}{N-1}
\end{aligned}
\end{equation}

\subsection{Fidelity Statistics for Mixed States}

In order to generalize this to realistic cases where there are unmeasured cogits which impact those being measured, which is equivalent to the case of mixed states in quantum information science, we must instead use the more general form of fidelity.

\begin{equation}
    \text{Fi}(\rho,\sigma) = \left(\text{tr}\sqrt{\sqrt{\rho}\sigma\sqrt{\rho}}\right)^2
\end{equation}

This $\sqrt{\rho}$ of a density matrix can be represented in series form as 

\begin{equation}
    \sqrt{\rho} = \sum_{n=0}^\infty (-1)^n \binom{\frac{1}{2}}{n}(I-\rho)^n 
\end{equation}

While it is more difficult to derive a clean analytic results for the probability distribution in this more general case in terms of the density matrices, it is straightforward to calculate computationally and then plug into the derivation in \ref{appendix_fidelity_statistics} above. 

 \subsection{Distance Statistics}\label{appendix_distance_statistics}

Returning to the Bures distance, the mean distance is thus

\begin{equation}
    \mathbb{E}[b(\rho,\sigma)] = \sqrt{1-\frac{1}{\sqrt{N}}}
\end{equation}

As $N$ grows large, the mean normalized distance between two random states rapidly approaches 1.

We note the reversed relationship between fidelity $f$ and our rescaled Bures distance $b$ -- that is, that higher fidelity means lower Bures distance, and vice versa, so we subtract the above from 1 to assess the probability of a certain Bures distance given a specific value for fidelity. The algebraic relation is $f=(1-b^2)^2 = 1-2b^2 + b^4$, and together these yield

\begin{equation}
\begin{aligned}
    \text{Pr}[b(\rho,\sigma)<v] = 1-\frac{(2v^2-v^4)^{N-1}}{N-1}
\end{aligned}
\end{equation}

So, for example, the probability of two states having a normalized distance of less than 0.95 is 

\begin{equation}
    Pr[b<0.95] = 1 - \frac{0.990494^{N-1}}{N-1}
\end{equation}

The exponential scaling means that even with relatively low-dimensional state vectors one quickly reaches the case where lesser differences are vanishingly unlikely. At $N=100$ there is only a $\sim 0.4\%$ chance of two state vectors having a distance of less than 0.95. At 500 cogits (or qubits, depending on the application) in the state vector, this probability decreases to 17 in a million. For a distance of 0.5 the probabilities are, respectively, $\sim 2.9\times 10^{-38}$ and $\sim 1.4\times 10^{-182}$. This gives the same rapid concentration of distance statistics required for effective use of ideas from hyperdimensional computing, as non-trivial proximities between uniformly random state vectors are so improbable as to indicate such proximities reflect meaningful information. 

\section{Security Metrics}

Perhaps the most directly intuitive metric of how well a hypothesis of cognitive state distribution corresponds to the ground truth.

The Jensen-Shannon distance (J), and specifically its quantum analog (QJS), provides a convenient metric, and is already normalized to values between 0 and 1.\cite{majtey_2005_quantum_jensen_shannon_divergence_distinguishability,Brit_2009_properties_of_quantum_jensen_shannon_divergence} It is a symmetric and smoothed version of quantum relative entropy, making it a natural selection of measure of distance between cogit HV. For a quantum state space this is in fact a true metric in a mathematical sense. \cite{Virosztek_2021_quantum_jensen_shannon_divergence_is_a_metric} While not a necessary feature for all potential mathematical considerations, is helpful for preemptively avoiding any potential issues which could otherwise arise from an informal metric in future work. Its classical analog is defined for distributions $P$ and $Q$, with mixture distribution $M = \frac{1}{2} (P+Q)$, as 

\begin{equation}
D_\text{J}(P,Q) = \sqrt{\frac{1}{2} \left( \sum_{x\in \chi} P(x) \log \left(\frac{P(x)}{M(x)}\right) + Q(x) \log \left(\frac{Q(x)}{M(x)}\right) \right)}
\end{equation}

 The quantum equivalent is defined in terms of density matrices for some density matrices in the same basis $\rho$ and $\sigma$, and mixture density matrix $\tau= \frac{1}{2}(\rho + \sigma)$, as 

\begin{equation}\label{equation_quantum_jensen_divergence}
    D_\text{QJS}(\rho, \sigma) = \sqrt{\frac{1}{2}\left( \text{Tr}(\rho(\ln \rho - \ln\tau)) + \text{Tr}(\sigma(\ln\sigma - \ln\tau)) \right)}
\end{equation}

A formulation of the QJS in terms of more direct probability values exists, and is more intuitive in this context. Since $\text{Tr}(A + B) = \text{Tr}(A) + \text{Tr}(B)$, and the trace of a matrix is the sum of its eigenvalues, we can rewrite this in terms of probabilities of different measurements. If we assume both $\rho$ and $\sigma$ represent pure states (i.e. we have measured all cogits with statistical relevance to the problem) or can be well-approximated as such, then it can be described based on the probabilities $p_i$ and $q_i$ of each possible measurement outcome from $\rho$ and $\sigma$ respectively. That is, $\ket{\psi} = \sum_i \sqrt{p_i} \ket{i}$ and $\ket{\varphi} = \sum_i \sqrt{q_i} \ket{i}$. In this case, the QJSD reduces to the classical JSD. 

\begin{equation}
    D_\text{QJS}(\rho, \sigma) = \sqrt{\frac{1}{2}\left( \sum_i p_i \log \left(\frac{2p_i}{p_i+q_i}\right) + q_i \log\left(\frac{2q_i}{p_i + q_i}\right)\right)}
\end{equation}

For the more general case with mixed states, coherences, etc., we must simply compute the series forms of $\ln(A)$ for each of the matrices in the QJSD, as seen in Equation \ref{equation_quantum_jensen_divergence} above. 

We can provide a nice upper bound on the $D_{QJS}$, however, as shown by Roga et al.\cite{Roga_2010_bounds_on_quantum_jensen_shannon_divergence} Formulating the bound in terms of our normalized Bures distance from Equation \ref{equation_normalized_bures_distance}, we have:

\begin{equation}
    D_{QJS} \leq \sqrt{-\left(\frac{1}{2}b(\rho,\sigma)^2\right)\ln\left(\frac{1}{2}b(\rho,\sigma)^2\right) - \left(1-\frac{1}{2}b(\rho,\sigma)^2\right)\ln\left(1-\frac{1}{2}b(\rho,\sigma)^2\right)}
\end{equation}

This bound is more useful for an attacker, since they are concerned with minimizing the QJSD, but may prove useful for defenders in some fashion we have not considered. A benefit of this bound is that we can use our earlier derivation of the statistical properties of Bures distances between random cogit HV to perform statistical testing, and then link that testing with the bound itself.

\end{document}